# FAST AND USER-FRIENDLY QUANTUM KEY DISTRIBUTION

Grégoire Ribordy, Jean-Daniel Gautier, Nicolas Gisin, Olivier Guinnard, Hugo Zbinden

Université de Genève, GAP – Optique, Rue de l'Ecole-de-Médecine 20, 1211 Genève 4
Email: gregoire.ribordy@physics.unige.ch

**Some guidelines for the comparison of different quantum key distribution experiments are proposed. An improved « plug & play » interferometric system allowing fast key exchange is then introduced. Self-alignment and compensation of birefringence remain. Original electronics implementing the BB84 protocol and allowing user-friendly operation is presented. Key creation with 0.1 photon per pulse at a rate of 486 Hz with a 5.4% QBER – corresponding to a net rate of 210Hz - over a 23 Km installed cable was performed.**

**Introduction**

The expansion of telecommunications during the past two decades has induced the need for the development of new cryptographic techniques offering high security as well as easy key management. The so-called public key cryptosystems, initially proposed by Diffie and Hellman [1] in 1975, were the first attempt to solve the key distribution problem. They make use of mutually inverse transformations for encrypting and decrypting. The encryption algorithm and key are made public – hence the name of public key cryptosystems - and the safety relies on the high complexity of the inverse transformation, unless the decryption – or private – key can be used. These ciphers form the class of asymmetric cryptosystems, of which RSA [2] is the most widely used. However, they suffer from a major flaw, namely the fact that their security relies on unproven assumptions about complexity theory. A fast algorithm could indeed break them. Although such an algorithm has already been proposed for quantum computers, it is not sure whether a similar one could also be found for classical computers.
The fact that this discovery would immediately jeopardise the secrecy of most electronic communications explains the strong interest in alternate encryption techniques. One could of course use symmetric ciphers. This class of cryptosystems includes Vernam's "one time pad", which is the only one offering proven absolute secrecy. In this case however, the problem is shifted to the distribution of the secret key safely between the emitter and the receiver. A spy could indeed intercept the key material and copy it. Quantum key distribution – often referred to as quantum cryptography – prevents this from happening. It complements symmetric ciphers, to simplify their implementation. This promising application has triggered a very strong interest in the industry as well as in the public. Its security is based on the principles of physics, and not on any unproven assumptions about complexity theory.

Quantum key distribution (QKD), first proposed in 1984 by Bennett and Brassard [3], allows two remote parties – usually called Alice and Bob - to generate a secret random key, used to carry out secure communications, without meeting or resorting to the services of a courier. It is based on the fact that, according to quantum physics, performing a measurement on an unknown state will in most cases disturb it. This property can be exploited to reveal an eavesdropper. Indeed, if the sequence of transmitted bits, encoded in non-orthogonal states, does not contain any errors, it can be inferred that no unlegitimate party tried to listen in. QKD was first demonstrated in 1989 by Bennett and his co-workers over 30 cm in air [4], and since then a long way has been covered by the quantum optics community. Various groups have indeed implemented it in optical fibres over distances up to tens of kilometres [5,6,7,8,9].

In this article, we would like first to consider the criteria that should be used to compare different experimental realisations of QKD. We then discuss an improved version of the "plug & play" QKD system, before looking at experimental results obtained. Finally, we discuss the respective advantages of various schemes, before concluding.

**Comparing Quantum Key Distribution Schemes**

Without going too much into details (see [10, 11] for a thorough description of the principle of QKD), quantum key distribution can be split in three elementary processes. First, Alice and Bob must generate a raw key, by exchanging photons over the quantum channel, which usually consists of an optical fibre. The term "raw key" comes from the fact that, due to experimental imperfections such as detector noise, the distributed bit sequence contains errors, even without tampering by an eavesdropper. Performing the distribution of this first key at a fast rate without introducing too many errors and complicated operation is the task of experimental physicists. The raw key creation rate ($R_{raw}$) and the quantum bit error rate (QBER) are the two quantities usually presented to characterise the performance of a QKD scheme.

The second step involves the estimate of the QBER within the raw key and the correction of these errors through discussion over a public channel. Performing this results in a shortening of the key to make up for the information revealed in the process to Eve. As the fraction of bits to discard increases with the QBER, it is essential to keep it as low as possible. This operation resorts to standard error correction procedures developed within the framework of classical information theory.

Finally, the third step reduces the information having leaked to Eve during the first two stages to an arbitrarily low value through a privacy amplification procedure. It also amounts to a reduction of the key length, which increases with the QBER. Indeed, the errors must be completely attributed to Eve. After this last procedure, Alice and Bob share exclusively a perfectly correlated sequence of bits, the useful key that can be exploited to carry out secure communications. Although not directly related to the physical QKD process, the last two steps do matter to experimental physicists. The way they are performed determines indeed the ratio of the length of the useful key to that of the raw key for a given QBER. They set an upper limit to the error rate, above which key distillation becomes impossible.

In order to predict the performance of a given QKD scheme the key distribution rate before error correction and privacy amplification $R_{raw}$ can be estimated from experimental parameters using the following formula (see [10] for a discussion of this expression):

$$R_{raw} = q \mu \nu \eta_t \eta_d \tag{1}$$

$q$ is a systematic factor depending on the chosen implementation. In the case of the BB84 four states protocol [3] for example, it is equal to ½, as half of the time the bases chosen randomly by Alice and Bob are not compatible. $\mu$ is the average number of photons per pulses. In a more rigorous treatment, one should replace this quantity by the probability to have a pulse containing at least one photon. Provided it is small, it is well approximated by the average of the Poisson distribution. $\nu$ the repetition frequency and $\eta_d$ the detector efficiency. Finally $\eta_t$ is the transfer efficiency between Alice's output and Bob's detectors, which can be expressed as:

$$\eta_t = 10^{\frac{-(L_f l + L_B)}{10}} \tag{2}$$

where $L_f$ corresponds to the attenuation in the fibre in dB/km, $l$ to the length of the link in km and $L_B$ to Bob's internal losses in dB.

The quantum bit error rate QBER, generally expressed as the relative fraction of wrong bits in the detected sequence, can be evaluated as the ratio of the probability of getting a false detection to the total probability of detection per pulse (see [10]):

$$QBER = \frac{\pi_{opt} \, p_{phot} + p_{noise}}{p_{phot} + 2 p_{noise}} \cong \frac{p_{noise}}{p_{phot}} + \pi_{opt}$$

$$= QBER_{det} + QBER_{opt} \tag{3}$$

with $p_{phot} = \mu \eta_t \eta_d$, we obtain:

$$QBER_{det} = \frac{p_{noise}}{\mu \eta_d \eta_t} \tag{4}$$

$p_{noise}$ and $p_{phot}$ are the probabilities to register a count arising from noise and from a photon respectively, while $p_{opt}$ is the probability for a photon to reach the wrong detector. The QBER consists of two parts. The first one, called QBER$_{det}$, is caused by the nois counts, mainly due to detector dark counts. The second part, QBER$_{opt}$, is caused by the propagation of photons to the wrong detector, because of incorrect phase or polarisation determination. When the transmission distance increases, the transfer efficiency $h_t$ becomes smaller, while $p_{noise}$ does not change. This results in an increase of QBER$_{det}$, which imposes the principal limitation on long distance QKD.

With the help of these equations, we can easily compute the raw key creation rate R$_{raw}$ and QBER under given experimental conditions. As the errors present in the raw bit sequence are suppressed during key distillation, the useful key creation rate R$_{useful}$ – after error correction and privacy amplification – constitutes the only figure of merit that actually matters. It is thus essential to determine to what extent R$_{raw}$ is reduced by the key distillation procedures, as a function of the QBER.
On the one hand, it is possible to use an estimate by Tancevski et al. [12]. The fraction of bits lost in the process of error correction as a function of the QBER is given for long strings (> 100 bits) by:

$$R_{ec} = \frac{7}{2} \text{QBER} - \text{QBER} \, log_2 \text{QBER}$$

(5)

This expression is valid for small QBER. $R_{ec}$ increases with the error rate, which implies that it is important to keep it low.
On the other hand, the fraction of bits lost through privacy amplification can be estimated by [11,13]:

$$R_{pa} = 1 + log_2\left((1 + 4\text{QBER} - 4\text{QBER}^2)/2\right)$$

(6)

It is assumed that all errors are caused by Eve's tampering and that she can gain a maximum information of 4 QBER / √2 [4] or about 2 QBER / ln2. This assumption implies that the average number of photon per pulse µ is small enough – typically 0.1 – to ensure that a low fraction of the pulses contains more than one photon. If this condition is not respected, the estimate does not apply anymore.

Finally, the useful key creation rate can be estimated:

$$R_{useful} = R_{raw}(1 - R_{ec})(1 - R_{pa})$$

(7)

When assessing different QKD schemes, two aspects must be considered. One must first look at the performance of the systems, in terms of key creation rate. Experimental results – in the form of measured $R_{raw}$ and QBER - for various schemes are available in numerous publications. They are difficult to compare though, because of the differences in experimental conditions. For example, the length of the transmission line, and hence the transfer efficiency, used for the tests is different in each case. Although most authors agree on using an average number of photon per pulse of 0.1 to ensure high secrecy, some use higher values, making comparisons impossible without scaling of the results. In our opinion, in spite of the fact that using values of $\mu$ larger than 0.1 may still be safe with present day technology, it should be avoided, if one wants to claim that the security is based on the law of quantum physics.

To make it possible to compare the key creation and error rates $R_{raw}$ and QBER for several QKD systems, one should scale these quantities using equations (1) and (4) for a given transmission distance and average number of photons per pulse of 0.1. One can expect to measure these scaled rates, if the experiment was repeated over this distance and with this average number of photons per pulse. If one were for example to use $\mu = 0.2$ in experimental tests, the scaled raw key rate would be smaller and the scaled QBER larger by a factor of approximately two. However, the problem is not completely solved yet. How should one proceed to compare systems with different scaled $R_{raw}$ and QBER? These two quantities depend on the relation between $\eta_d$ and $p_{noise}$, which varies from one photon counting detector system to the other. The solution is simply to compute the useful key creation rate estimate from the scaled $R_{raw}$ and QBER using equations (5) to (7). It appears now clearly why it is very misleading to consider QKD systems with average photon number larger than 0.1. When scaling the results to $\mu = 0.1$, $R_{raw}$ is not only decreased, but the QBER is also increased. This implies that when computing the useful rate $R_{useful}$, it will be further reduced by this extra QBER contribution. In this case, it is important to remember that because the estimate of bits lost during privacy amplification is not valid if $\mu$ is larger than 0.1, one must first scale the results before evaluating the useful rate.

The second aspect to consider when comparing systems is the intrinsic simplicity of use. What adjustments are required when connecting a QKD system to a transmission line? How difficult and frequent these adjustments must be? All systems using gated detectors require for example high precision (nanosecond or better) synchronisation of Alice and Bob, in order to trigger the photon counters just before the arrival of a photon. This adjustment however does not need to be repeated too frequently. Indeed it depends mainly on the length of the transmission line, which varies only slightly. Other adjustments include polarisation or interferometer alignment.

**Improved Version of the "Plug & Play" QKD System**

Quantum key distribution has been demonstrated in optical fibres by several groups using either polarisation encoding schemes [5, 6, 7], or phase encoding ones [8, 9]. The main difficulty with polarisation encoding is the need to keep the polarisation stable over distances of tens of kilometres. It was indeed shown that, because of the birefringence of the fibres and the effect of the environment, the output polarisation state fluctuates randomly. Although these variations are slow enough to allow active polarisation tracking, such a compensation could be unpractical in a real system.

To avoid this problem, an alternative is to encode the value of the bits in the phase of the photons instead of their polarisation. One must then use an interferometric key distribution system, consisting usually of two Mach-Zehnder interferometers in series separated by the transmission line. However if one wants to keep $QBER_{opt}$ low, high interference visibility must be guaranteed. Alice and Bob must make sure that both of their interferometers are identical: same coupling ratios and same path length. They must keep them stable within a few tens of nanometer during the course of the transmission, which constitutes the major difficulty. Moreover such a system still requires polarisation control in the interferometers. The polarisation of both interfering contributions must be parallel, to ensure high interference visibility.

We introduced in 1997 a new phase encoding quantum cryptography system intrinsically stable and polarisation independent [14]. This interferometer does not require path alignment thanks to time multiplexing. In addition polarisation control is made superfluous by the use of Faraday mirrors to cancel the effect optical fibre birefringence. Apart from synchronisation of Alice and Bob, no adjustment at all is necessary. This is why we refer informally to this scheme as "plug & play". It was tested and showed high interference visibility and outstanding stability. A useful key creation rate of approximately 0.5 Hz with the B92 protocol [15] was achieved [16]. In order to assess its potential, we decided to introduce an improved version allowing higher key creation rate and implementing the safer four states BB84 protocol. Moreover, we also decided to try to use photon counters with simpler cooling requirements, and to develop original electronics allowing automated operation of Alice and Bob.

As it induces several problems, increasing the key distribution rate does not simply amount to turning up the repetition frequency, but requires re-engineering of the whole system. To understand the first problem, it is essential to remember that in the "plug & play" system, laser pulses are not emitted by Alice towards Bob. They are emitted by Bob, travel to Alice where they are reflected before coming back to Bob for detection. Because of the Michelson interferometer configuration of the 97 implementation, an important fraction of the light pulses is directly reflected straight into the photon counting detector. It can increase the error rate through two mechanisms. First, if this light reaches the detector at the same time as a photon coming back from Alice, it can generate a false count due to the high intensity difference, and increase the error rate. Choosing the lengths of the various optical fibres so that the light reflected within Bob's interferometer does not impinge on the detector when it is gated could in principle solve this problem. This is however not sufficient. We have indeed measured the detection efficiency for InGaAs/InP avalanche photodiodes (APD) in photon counting regime for pulses impinging before gating. Results are shown in Figure 1 for two different detector temperatures. One notices that when the light pulses reach the detector when it is gated, the efficiency is about 10%.

At 77 K we observe that the detection efficiency for pulses arriving before the gate first decreases by two orders of magnitude during the first 20 ns. At longer time, the slope becomes very small. Although the remaining efficiency is low, it may cause problems with strong reflections containing thousands of photons. This implies that, contrary to what is usually assumed, avalanche photodiodes in photon counting regime are not really blind when not gated. At 173 K, the decrease of the efficiency is faster, and thus the timing discrimination of the detector better. This problem could not be observed in the 97 system. The repetition rate of 1 kHz was indeed low enough to ensure that there was never more than one pulse in the optical fibre. The round trip time of flight of a pulse was of approximately 230 μs for the 23 km long line while the time interval between the emission of subsequent pulses was 1 ms. To get round this difficulty, we developed a modified version of the "plug & play" system shown in Figure 2. It does not feature any strong reflections of emitted pulses into the photon counters. Although optical fibre connectors or components may reflect some of the light, this intensity is low and does not constitute a problem. Bob injects a light pulse through the circulator C, which splits at coupler C1. The first half pulse travels through the short arm to the polarising beamsplitter PBS. Polarisation evolution in this arm is set with a polarisation controller (three loops on the diagram) so that the pulse is completely transmitted at the PBS. It then propagates to Alice where it splits again at coupler C2 to provide a timing signal upon detection by the linear detector $D_A$. The pulse then travels through Alice's equipment and is reflected back to Bob. Thanks to the Faraday mirror, the birefringence of the optical link is compensated, and the pulse comes back orthogonally polarised, which implies that it is now reflected by the PBS and takes the long arm. Bob finally applies a phase shift $\phi_B$ with its modulator $PM_B$. The second pulse propagates through both arms of Bob's system in reverse order: long arm first, reflection by the PBS, Alice and back, transmission by the PBS, short arm. Alice applies to it a phase shift $\phi_A$. When applying their phase pulses, Alice and Bob must be sure to affect only the intended pulse and not both. This condition is most stringent for Alice where the pulses follow each other, separated by a time interval corresponding to the length difference of the short and long arms – typically 200 ns. Since both pulses travel exactly the same optical path, they reach the coupler C1 simultaneously with identical polarisation, giving rise to interference. According to the phase applied by Alice and Bob, the pulse chooses either deterministically or probabilistically a detector. At Alice's, the attenuator A is set so that the average number of photon per pulse pair μ is 0.1 when leaving her system. In contrast to the 97 setup, polarisation alignment is required in Bob's system. The polarisation evolution is set in each arm to maximise intensity at the output port of the PBS. In addition, due to high polarisation dependent losses of Bob's phase modulator, a second polarisation controller is necessary in the long arm. In an industrial realisation, polarisation-maintaining fibres would, of course, be used. When switching from constructive to destructive interference, we measured with classical light an extinction ratio of 28.6 ± 0.8 dB. The polarisation separation of the PBS, being of approximately 30 dB, constitutes the limiting factor for interference visibility. This extinction does not vary with time. It does not depend either on the optical fibre connected between Alice and Bob. A second advantage of this system over the 97 "plug & play" setup is the fact that two detectors can be easily connected to the output ports of the interference coupler allowing simple implementation of the BB84 protocol. One should note that Alice must also use the detector $D_A$ to monitor the incoming light intensity in order to prevent Eve from obtaining the value of her phase shift by sending strong pulses into the system and measuring their phase.

A second difficulty related to high repetition rates is Rayleigh backscattering. The light travelling in an optical fibre undergoes scattering by inhomogeneities. A small fraction of this light is recaptured by the fibre in the backward direction. Due to the intrinsic bi-directional nature of the "plug & play" setups, pulses travelling to and back from Alice must intersect at some point in the line. Their intensity is however strongly different. The pulses coming from Bob are more than 30 dB brighter than the ones coming back from Alice and containing less than 0.1 photon on average. Backscattered photons can accompany a pulse propagating back to Bob and induce false counts. Although the backscattering probability is very low, the high intensity ratio between forward and backward propagating light and the single photon sensitivity of the detectors can yield increased QBER. To prevent this problem, we introduced a storage line SL in Alice's system. Bob emits his pulses in the form of trains that travel to Alice and fill the storage line. He waits until a train comes back, detects the pulses, and sends the next one. Intersection of forward and backward travelling light takes place inside the storage line and not in the transmission line. As it is located behind the attenuator, the intensity ratio is low enough so that the backscattered intensity cannot cause false detections. The losses induced by this storage line do not amount to a reduction of the transfer efficiency, because the attenuator will be set to keep the average number of photons per pulse at 0.1 at the output of Alice. However the effective repetition frequency of the system is reduced in such a configuration. The emission and detection of the pulses are sequential and separated by the time taken by the photons to travel through the transmission line, the storage line, and back. A storage line half as long as the transmission line amounts to a reduction of the effective repetition frequency by a factor of approximately three. Increasing the length of the storage line can reduce this effect. If it is too long though, its losses may start to matter, as it becomes difficult for Bob to send subnanosecond pulses still featuring an average of 0.1 photons per pulse at the output of Alice.

At high repetition frequencies, detectors constitute a third problem. When implementing a long distance QKD system, it is essential to choose the wavelength of the photons within the second or third telecom window – around 1310 and 1550 nm respectively – in order to minimise the transmission losses and maximise the transfer efficiency. Detecting single photons at these wavelengths with a reasonable level of background noise is however difficult. One usually uses Germanium avalanche photodiodes (Ge APD) biased above breakdown in the so-called Geiger regime. However at room temperature, these detectors show high dark counts levels. They must then be cooled to the liquid nitrogen boiling temperature - 77 K - to be used for QKD. When an avalanche arises in such a diode, a current flows through the junction and some of the charges get trapped by defects. The subsequent release of these carriers can trigger avalanches – called afterpulses - causing false detections, and thus increase the QBER. This effect becomes particularly significant when the separation between the pulses is small because of high repetition frequency. To get round this problem, we turned to InGaAs/InP avalanche photodiodes (Fujitsu FPP5W1KS). They are operated in gated mode with 2ns activation. This prevents the current from flowing too long, and limits the population of the trapping levels. Under similar conditions, these detectors show lower dark count rates than Ge APD's. It is thus possible to operate them at higher temperature, which in turns decreases the lifetime of trapped charges. A temperature of 173 K offers a trade-off between dark counts and afterpulses. This temperature is obtained by heating a cryostat immersed in liquid nitrogen. Operating detectors above the temperature of liquid nitrogen also offers the advantage of making cooling easier. This temperature does not lie too far from the lower limit of Peltier cooling, which would be really interesting in the event of a real world application. Figure 2 shows the performance of these detectors. The probability of getting a noise count $p_{noise}$ is plotted versus $h_d$ the detection efficiency. We typically use them with $p_{noise} = 10^{-5}$ and $h_d = 10\%$.

Finally, the driving electronics had to be redesigned to allow repetition at 2.5 MHz, as well as train generation and detection. Bob's electronic circuit contains an oscillator beating at 10 MHz. The various triggering signals are then generated by counting clock strokes. Fine adjustment between two such strokes is performed through delay lines with an accuracy of 100 ps. The 260 ps wide optical pulses are generated with a Fujitsu DFB semiconductor laser at 1310 nm and are separated within a train by 400 ns. Approximately 100 ns before gating the detectors, the UTP annealed proton exchange $LiNbO_3$ phase modulator is triggered. The applied phase shift is determined by the value of a random bit generated in real time by a specially developed noise source (NS). For each detector count, the index of the train and of the pulse, the outcomes of both detectors and the value of the random phase shift are stored in a buffer. A PC, where the data are downloaded and processed, controls the board. As discussed above, the "plug & play" setups do not require any stabilisation or active adjustment of either the interferometer path length or the polarisation. However, they do require, as all other systems using gated photon counters, precise timing of the detectors activation. Basically, we have to perform a measurement of the total round trip time of flight of a pulse – approximately 300 µs with the storage line - with a precision of the order of 100 ps. We have thus implemented a photon counting optical time domain reflectometer mode within our system. It measures the time interval between the emission and the detection of a macroscopic pulse in two stages. To initiate the first stage, the operator is asked to enter the length of the transmission line with a precision of plus or minus five kilometres. The system scans this range by steps of 1 ns to find the reflection of Alice's mirror, which is orders of magnitude stronger than all the other parasite reflections in the system. This takes approximately 2 minutes. In the second stage, the system then scans a 2ns wide region surrounding the detected reflection by steps of 100 ps to fine-tune the timing.

This takes about one minute. This procedure is carried out automatically by the computer. Hence the system can be ready to perform QKD in less than five minutes after the connection of an optical fibre line.

Alice's system is simpler. A PIN photodiode $D_A$ with fast amplifier is used to detect the pulses and generate a timing signal. The signal is delayed to let the pulse propagate through the storage line before triggering Alice's electronic board. This circuit then takes two bits – encoding one of four states - and passes them to the phase modulator controller. These two bits are downloaded from a file to make comparison of Alice's and Bob's keys simpler, but they could also be generated by a noise source. A 100ns wide voltage step is accordingly applied to the Ti indiffused $LiNbO_3$ device. The pulses cross the modulator twice. The phase is thus modulated on a round trip. The modulator must feature low polarisation dependent losses, if the Faraday mirror is to compensate the birefringence of the line. Alice's timing accuracy is far less stringent than Bob's. She has to be able to apply her phase shift on one of the two half pulses separated by 200 ns. A precision of about 50ns is then sufficient. Moreover, her delay does not depend on the length of the line, but only on the time interval between the detection at $D_A$ and the application of the phase shift. They can then be set once for all in the laboratory.

**Experimental results and comparisons with other systems**

This improved setup was tested in the laboratory on a 4.9km optical fibre spool featuring losses of 4.6 dB, as well as on a 22.8km optical fibre running under Lake Geneva and connecting the Swiss cities of Geneva and Nyon. This line features losses of 10.5 dB. The results are presented in Figure 4 and summarised in Table 1. The points show the raw key creation rate $R_{raw}$ versus the quantum bit error rate QBER for both transmission fibres. The solid lines show a linear fit of the experimental data, while the dotted lines represent the estimate of the useful rate $R_{useful}$ after error correction and privacy amplification for the fitted values of $R_{raw}$. It is computed with equations (6) and (7). These results were obtained with $\mu = 0.1$. The experimental parameter is the gate voltage applied on the detectors, which controls the detection efficiency ranging from 2.5% to 20%, and the dark count probability. Key distribution was performed at a raw rate of 1630 and 486 Hz for 4.9 and 23km respectively with QBER of 4 % and 5.4 %. To verify the quality of the theoretical predictions, we can note that the measured interference visibility yields a value of 0.14% for $QBER_{opt}$. On the other hand, $QBER_{det}$ can be estimated to be 5.1% for the 23 km link. Afterpulses and Rayleigh backscattering should not contribute in any noticeable way to the QBER. These contributions sum up to 5.2%, which is close to the measured QBER of 5.4%. These results translate to maximum estimated useful rates of 870 Hz and 210 Hz for the laboratory and field measurement respectively. One notices that the useful rate increases with the QBER. However it would eventually reach a maximum before decreasing. In both cases, the obtained raw rates yield useful rates close to the highest point on the curve.

In Table 2, we show some results that confirm the effect of the parameter $\mu$.

We performed key distribution with various setting of Alice's attenuator corresponding to average photon numbers of 0.1, 0.2 and 1 on the 23 km cable and with gate voltage such that $\mathbf{h}_d$ = 11% and $p_{noise}$ = $10^{-5}$. These results illustrate well the fact that when µ is increased by a factor two, the QBER is divided and $R_{raw}$ multiplied accordingly. The same thing happens with µ going from 0.1 to 1. The QBER remains larger than expected though. This probably comes from the fact that $QBER_{det}$ only is affected this parameter. At low error rate, $QBER_{opt}$ starts to be noticeable.

We also justified the use of the storage line, by demonstrating an increase of the error rate caused by Rayleigh backscattering. We studied the value of the QBER as a function of the number of pulses per train. The results are shown in Figure 5. The QBER is constant as long as the length of the trains is smaller than twice the length of the storage line. When the trains' length is increased beyond the storage capacity, the QBER increases. Overfilling by twenty pulses is sufficient to double the QBER.

The operating temperature of the detectors was also varied. At 193K and 203K, no significant change was observed. At 213K however, performance clearly deteriorates, through increased QBER. Figure 6 shows the raw key creation rates versus the QBER for these temperatures. These results are better than expected. When plugging the values of $\mathbf{h}_d$ and $p_{noise}$ obtained by characterising the detectors in equations (1), (3) and (4), the dark counts levels appear too high to perform QKD with reasonable error rate above 173 K. The fact that QKD is actually possible indicates that the effective $p_{noise}$ must be lower than the value obtained during characterisation. A possible explanation is that when the detectors generate counts, their noise probability is reduced. All in all, these results suggest that the possibility to operate these APD's for QKD under Peltier cooling is not far.

Finally, it is interesting to compare these results with those obtained by other groups. We will restrict ourselves to the experiments carried out by the two groups that have implemented QKD in optical fibres over distances of more than 20km and with photons at 1310 nm, and published results to date. First Townsend [8] of BT performed QKD in 1998 over distances ranging from 5 km to 55 km using a standard phase encoding system. For 25 km, he obtained a raw key creation rate of 500 Hz with a quantum error rate of 2%. However it is essential to note that he used an average number of photon per pulse of 0.15, above the commonly used value of 0.1. This prevents us from directly estimating the useful rate. We must first scale these results to what could have been expected, if µ had been set to 0.1. The scaled raw rate is 360 Hz and the scaled QBER 3 %, yielding a useful rate of 220 Hz.
Second, Hughes et al. [9] of Los Alamos National Laboratory performed in 1997 an experiment over 24 km also with a standard phase encoding system and obtained a raw rate of 20 Hz and a QBER of 1.6%. In this case too, the average number of photon per pulse was chosen at 0.4, well above the commonly used value of 0.1. When scaling these results, one obtains a raw rate of 6 Hz and a QBER of 6.4 %, yielding a useful rate of 2.2 Hz. Both these examples illustrate how a high µ can boost the performance of a QKD system at the expense of secrecy.

As these QKD systems are very similar, the difference between the results obtained by BT and Los Alamos can be explained by the fact that they used a repetition frequency of respectively 1 MHz and 10 kHz. There is no reason to think that Los Alamos could not also increase its repetition frequency to achieve similar key rates, if their detectors do not yield excessive noise through afterpulses.

Due to the necessary use of a storage line and the waiting delays it induces, the "plug & play" setup is intrinsically slower than standard systems at a given repetition frequency. However this rate reduction can be kept low by using a suitably long storage line, and is compensated by the truly superior self-alignment and stability properties.

**Conclusion**

After the discussion of some guidelines for comparison of different QKD experiments, we introduce an improved version of the "plug & play" system. We present then experimental results of key creation over 4.9 km and 23 km at useful rates of respectively 870 Hz and 210 Hz. Original electronics allowing easy operation of the system is described. It appears clearly that the main experimental difficulty remains the photon counting detectors. Their dark counts constitute the main contribution to QBER and limit thus the transmission distance. Moreover, simpler cooling – possibly using Peltier elements – should be considered in the view of an industrial application if compatible with noise requirements. The reduction of afterpulses by reduction of the trapping defects density is also important for further increase of the repetition frequency. However these photon counting APD's can only be improved by manufacturers, which unfortunately seem to lack interest.

The performance of the improved "plug & play" system is then measured up other comparable systems. The useful key creation rate obtained is similar to the best one to date in similar conditions. We have shown that the problems caused by the operation of the "plug & play" systems at fast repetition rate can be overcome to yield high performance, while maintaining unbeatable adjustment simplicity. In addition, "plug & play" systems may well constitute the only possibility for QKD over highly unstable optical fibres, such as those present in aerial cables.

**Acknowledgment**

We would like to thank Swisscom for placing two optical fibres of the Geneva – Nyon cable at our disposal for the duration of the experimental tests, and Mario Pasquali for his programming work. We greatly benefited from stimulating discussions with Bruno Huttner, and our colleagues of the "Physics of Quantum Information" TMR network. This work was partially supported by the Esprit project 28139 through the Swiss OFES.

**Tables**

**Table 1:** Experimental performance of the improved "plug & play" quantum key distribution system tested over an optical fibre spool in the laboratory and an installed cable.

| Environment | Distance [km] | Average number of photons μ | $R_{raw}$ [Hz] | Measured QBER | Estimated $R_{useful}$ [Hz] |
|---|---|---|---|---|---|
| Laboratory | 4.9 | 0.1 | 1630 | 4 % | 870 |
| Installed cable | 22.8 | 0.1 | 486 | 5.4 % | 210 |

**Table 2:** Illustration of the influence of the average number of photons per pulse μ.

| Distance [km] | Average number of photons μ | $R_{raw}$ [Hz] | Measured QBER |
|---|---|---|---|
| 22.8 | 0.1 | 129 | 1.51 % |
| 22.8 | 0.2 | 265 | 0.77 % |
| 22.8 | 1 | 1462 | 0.2% |

**Table 3:** Comparison of experimental realisations of quantum key distribution.

| Group | Distance [km] | μ | $R_{raw}$ [Hz] | Measured QBER | Scaled $R_{raw}$ [Hz] | Scaled QBER | Estimated $R_{useful}$ [Hz] |
|---|---|---|---|---|---|---|---|
| Geneva | 22.8 | 0.1 | 486 | 4.5 % | NA | - | 210 |
| BT | 25 | 0.15 | 500 | 2 % | 360 | 3 % | 220 |
| Los Alamos | 24 | 0.4 | 20 | 1.6 % | 6.1 | 6.4 % | 2.2 |

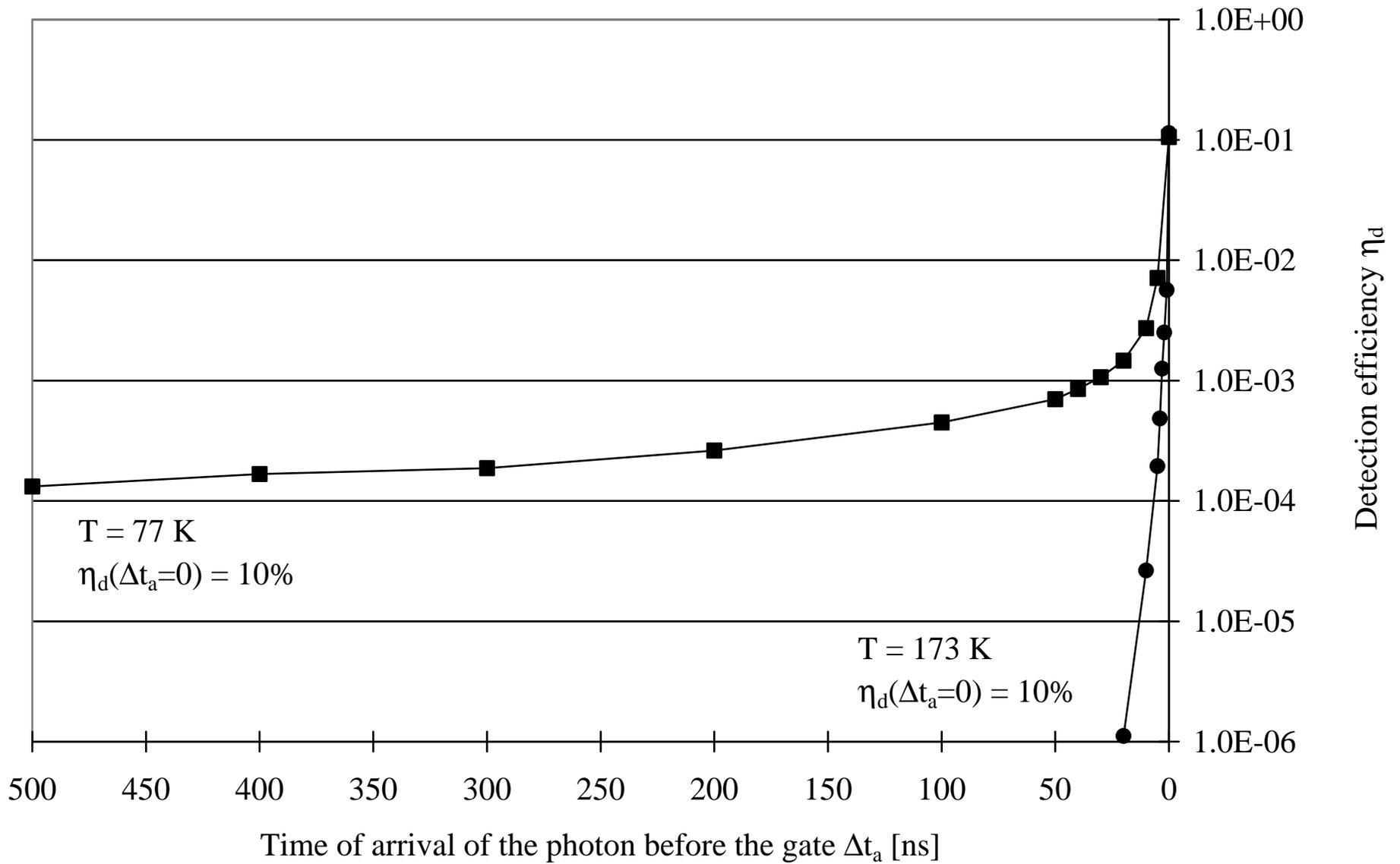

Figure 1 : InGaAs/InP avalanche photodiodes quantum detection efficiency against the time difference between the arrival of the photon and the activation of the detector.

Figure 2 : Schematic diagram of the improved "plug & play" quantum key distribution system.

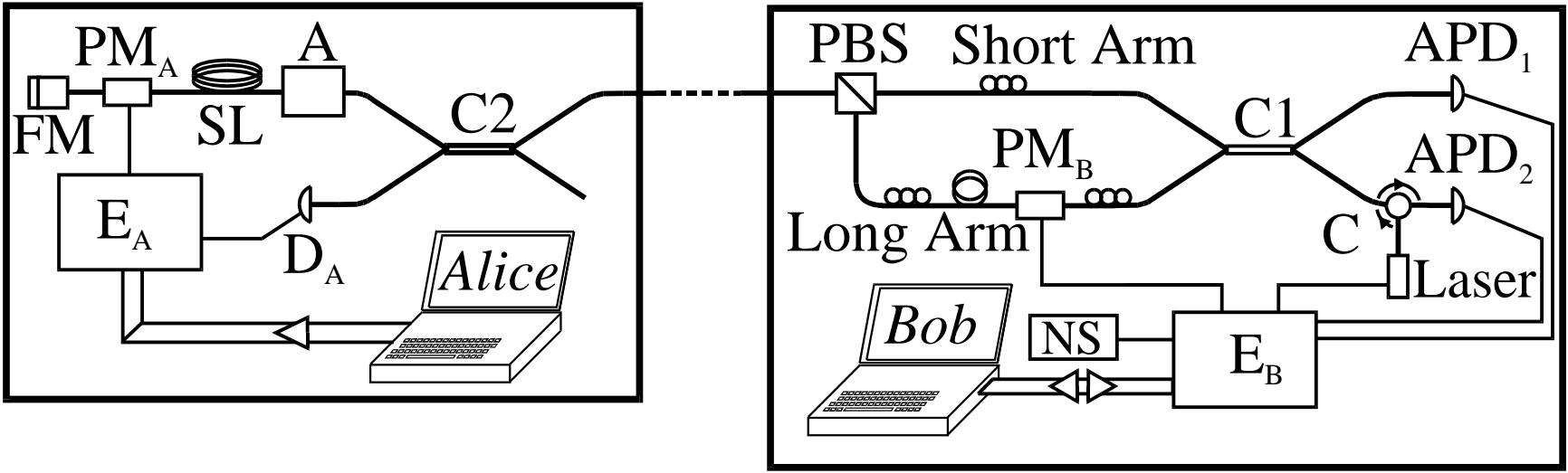

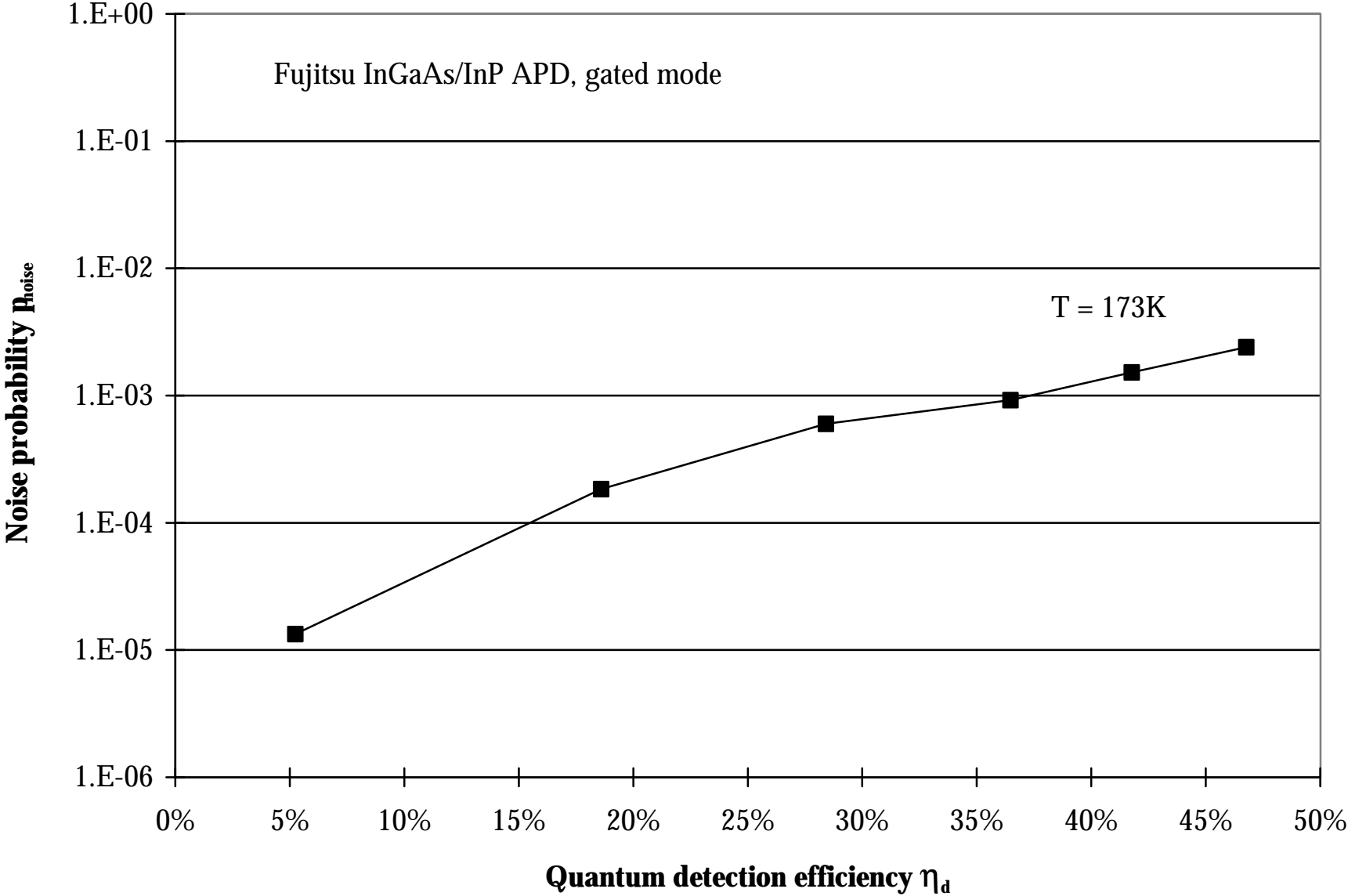

Figure 3 : Typical performance of InGaAs/InP avalanche photodiodes in photon counting regime at 173 K. Noise count probability against quantum detection efficiency.

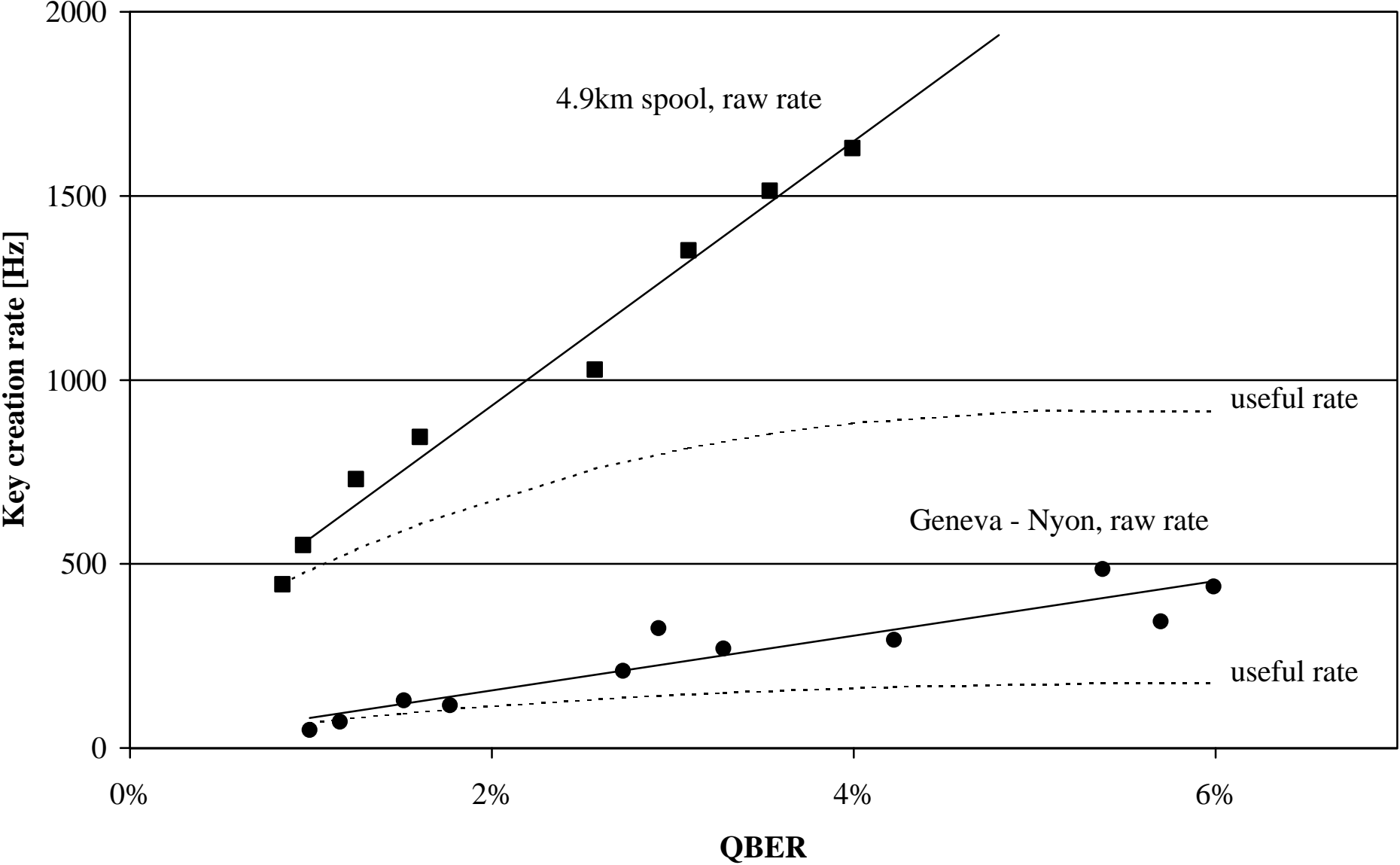

Figure 4 : Raw and useful key creation rates against QBER for transmission along 4.9km spool and 23km installed cable.

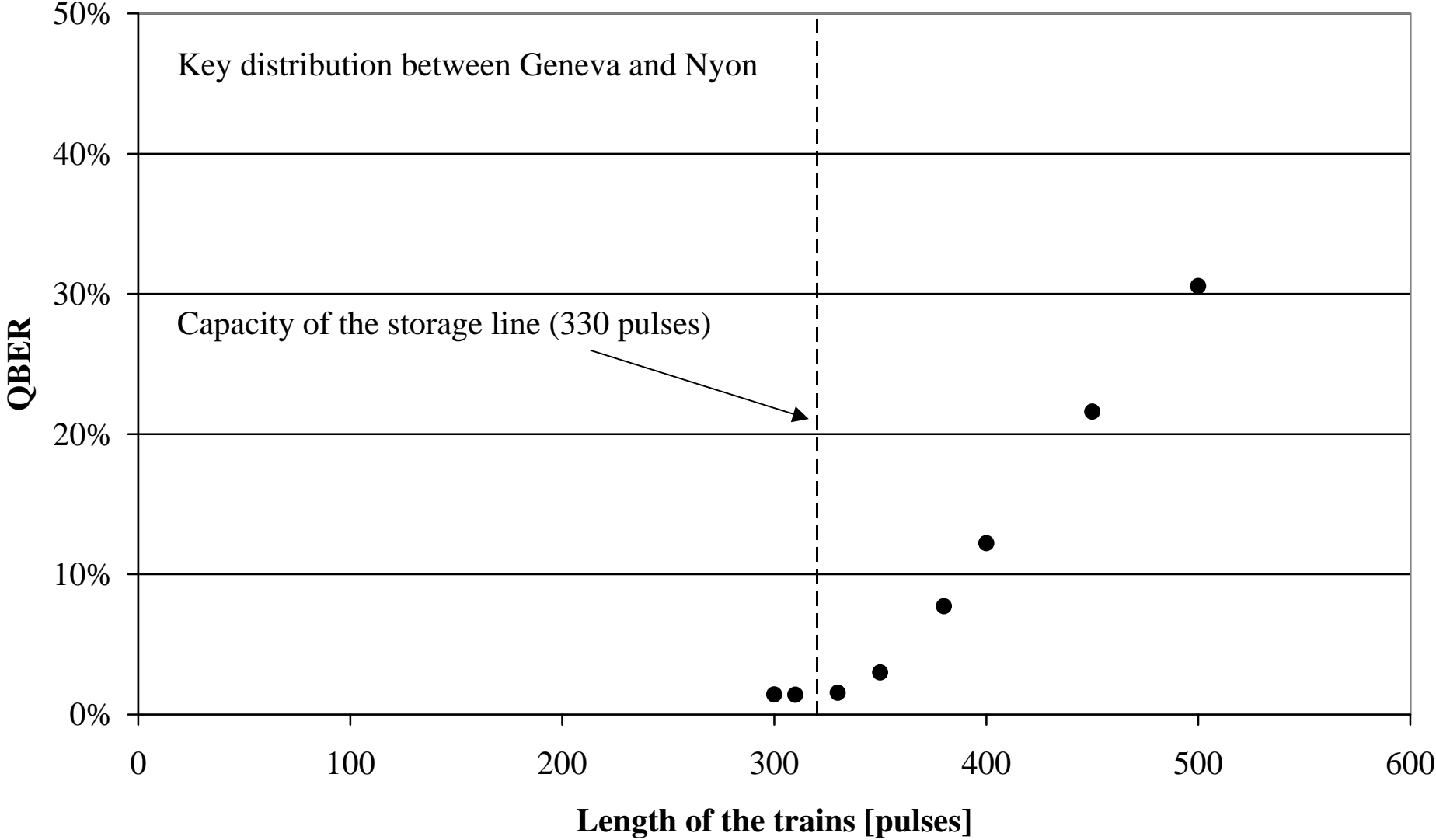

Figure 5 : Quantum bit error rate against length of pulse trains for transmission along 23km installed cable.

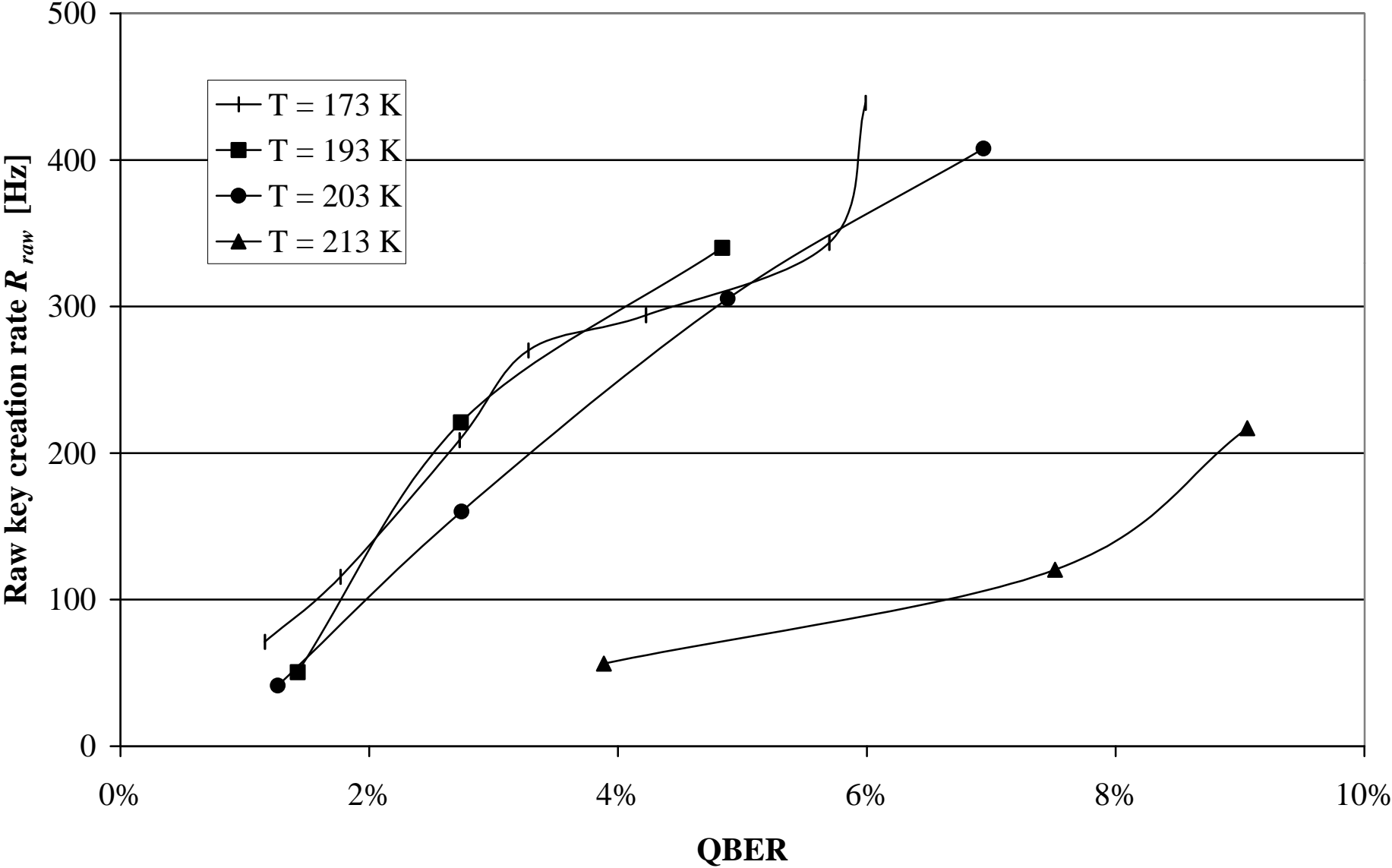

Figure 6 : Raw key creation rate against QBER for different detector operating temperatures. Transmission along 23km installed cable.